\documentclass[referee]{aa}
\usepackage{graphicx}
\input psfig.tex

\def\lsim{\, \lower2truept\hbox{${<
\atop\hbox{\raise4truept\hbox{$\sim$}}}$}\,}
\def\gsim{\, \lower2truept\hbox{${>
\atop\hbox{\raise4truept\hbox{$\sim$}}}$}\,}

\begin{document}

\title{$\Lambda$CDM and the Distribution of Dark Matter \\  in Galaxies: 
a  Constant--Density Halo around \\ DDO 47}

\author{ P. Salucci \inst{1}
\and F. Walter \inst{2}
\and A. Borriello \inst{1} }

\offprints{P. Salucci}

\institute{International School for Advanced Studies SISSA/ISAS, Trieste, I \\
\email{salucci@sissa.it, borri@sissa.it}
\and
California Institute of Technology, Pasadena, CA 91125, USA \\
\email{fw@astro.caltech.edu}
}

\titlerunning{Constant--Density Halo around DDO 47}

\date{Received .... accepted .....}

\abstract{

In this paper we present a test case for the existence of a core in the density distribution of 
dark halos around galaxies. DDO 47 has a rotation curve that increases linearly from the first 
data point, at $300 \ pc$, up to the last one, at $ 5 \ kpc$. This profile implies the presence 
of a (dark) 
halo with an (approximately) constant density over the region mapped by data. This evidences the
inability of standard $\Lambda$ Cold Dark Matter scenario to account for the dark matter 
distribution around galaxies, and points toward the existence of an intriguing
halo scale--length of homogeneity. This work adds up to the results
of Blais-Ouellette et al (2002), Trott \& Webster (2002), Binney \& Evans (2002), de Blok \& 
Bosma (2002) and  Bottema (2002) in suggesting that 
CDM theory should incorporate, as an intrinsic 
property at galactic scales, a ``density 
core" feature.}

\maketitle
%%

%________________________________________________________________
%%
\section{Introduction}

Rotation curves (RC's) of disk galaxies are the best probes for dark matter (DM) on galactic 
scale. Although much progress has been made over the past 20 years, it is only very recently that 
we start to shed light on crucial aspects of the DM {\it distribution}. Initially, the main focus 
was on the presence of a dark component; this later shifted to investigating the ratio of dark to 
visible matter (Salucci and Persic, 1997).
Today, the focus is mainly on the actual density profile of dark halos (e.g. Salucci, 2001) A 
cored 
distribution, i.e. a  density profile flat out to a radius 
that is a significant part of the disk size,  has been often adopted (e.g. Carignan
\& Freeman, 1985), although the implications of this distribution appeared only 
after that  cosmological $N$--body simulations found that Cold Dark Matter (CDM) virialized halos 
achieve a cuspy density profile
(Navarro, Frenk \& White, 1995, hereafter NFW):
\begin{equation}
\rho_{CDM}(r) =\frac{\rho_s}{x(1+x)^2}
\label{rhoCDM}
\end{equation}
where $x= r/r_s$, $r_s$ and $\rho_s$ are the characteristic inner radius and density and the 
simulations' spatial resolution has  recently  reached  $1/10 r_s$. The halo 
virial radius $R_{vir }$ is the radius within which the mean halo density is $\Delta_{vir}(z) $ 
times the mean cosmic density at that red shift (see Bullock et al.,
2001). The virial mass $M_{vir}$ and the
corresponding virial velocity are related by: $V_{ vir}^2 \equiv G M_{vir} /
R_{vir}$. The concentration parameter is defined by $c \equiv
R_{vir}/r_s$. With the above definitions, the NFW circular velocity can be written as:

\begin{equation}
V_{CDM}^2(r)= V_{vir}^2 
\frac{c}{A(c)} \frac {A(x)}{x}
\label{Vh}
\end{equation}
where $A(x)= \ln (1+x) - x/(1+x)$.
Numerical simulations show that in  CDM halos  the virial mass, the 
virial radius and the concentration are  mutually related:
 objects considered at $z=0$ within the 
cosmological scenario with $\Lambda=0.7$, $\Omega_0=0.3$ and $h=0.7$ have:

\begin{equation}
c \simeq 21 \ \left ( \frac{M_{vir}}{10^{11}\ M_{\odot}}\right )^{-0.13} 
\\
R_{vir} \simeq 120 \left ( \frac{M_{vir}}{10^{11}\ M_{\odot}}\right )^{1/3} \ kpc  
\end{equation}

(Wechsler et al., 2002). By applying the above to a reference mass of $5\times 10^{10} M_\odot$, 
reasonable for DDO 
47, we get: $c \simeq 22$ and $r_s \simeq 4 \ kpc$, that are very solid guesses in view of the 
weak mass dependence of these quantities. The available HI data, then, map, exactly the 
region in which a NFW halo changes its velocity slope from 0.5 to 0. The relationship (3) frames 
the $\Lambda$CDM  halo around DDO 47,  but will not enter in the 
crucial evidence we bring for $\Lambda$CDM;    the evidence, in fact,   will concern the theory at 
the most  fundamental level of eq (2).  

Discrepancies between the universal profile of CDM
and the mass distribution of the dark halo as inferred from the RC has
emerged a few years ago (Moore, 1994)
At the present, the existence of a crisis for CDM is seriously considered  (Blais-Ouellette et al 
(2002), Trott \& Webster (2002), Binney \& Evans (2002), de Blok \& 
Bosma (2002), Bottema (2002)), but the  opposite  
view is also claimed (van den Bosch et al., 2000, Primack, 2002) and   so as  the view for which   
 
NFW halos fare badly but not worse than other profiles (Jimenez et al., 2002).

Our  previous works approached the issue in two 
complementary ways: {\it i)} we derived the accurate mass structure of
halos around galaxies and then tried to fit them inside the CDM scenario
(Borriello \& Salucci, 2001)  and {\it ii)} we tested strategic
CDM features by means of appropriate available kinematical data (Salucci, 2001).

The recent HI data (Walter \& Brinks,
2001, hereafter WB01) and $I$--band surface brightness photometry
(Makarova et al., 2002) relative to  the dwarf galaxy DDO 47 give the 
opportunity to combine these two  approaches,  provide also an exemplar test case for CDM.  In 
fact: 
{\it a)} the RC of DDO 47 extends out to $\sim 9$ disk scale-lengths, which correspond to 1.3 NFW 
halo scale--length $ r_s$, at a spatial resolution of $1/7 r_s$ (i.e.  one 
disk scale-length)
{\it b)} the HI disk surface density  decreases sharply with radius: 
its flat contribution to the circular velocity does not mimic the solid-body profile of a 
constant density halo, and therefore does not complicate the mass modeling. 
{\it c)} the 
galaxy is of low luminosity: the content of luminous mass is small with respect to the dark one 
(e.g. Persic and Salucci,  1988) and consequently easy to take into account.

With these favorable circumstances, we are able to correctly investigate the mass structure of 
the dark halo around DDO 47, and eventually to discover the inner $r^{-1}$ signature of the NFW 
universal profile.  Finally,  we do not consider the compression exerted by  baryons when they 
infall on dark halos, in that this process makes CDM halos density profiles even
steeper than the original NFW ones (Blumenthal, 1986). 

DDO 47 is assumed at a 
distance of 4 Mpc, so that $1"=19.38 pc$ . The crucial results of this paper do not depend on the  
actual value for the galaxy distance, however, we will discuss the marginal role it plays. 
 
%%
%c
\begin{figure*}
\center{\psfig{file=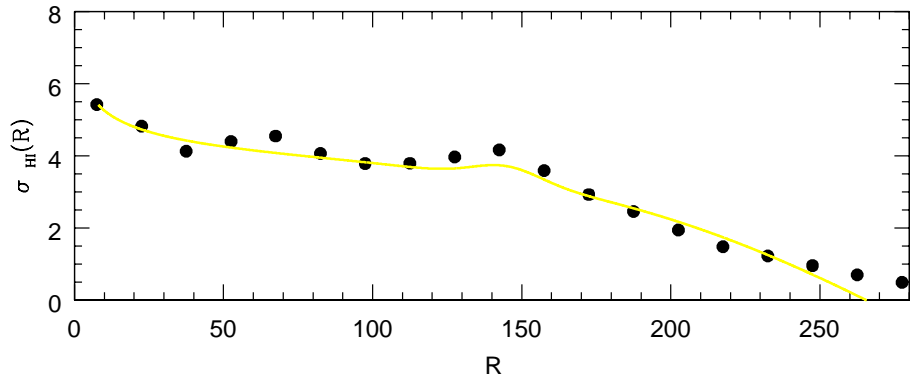,width=140mm,height=140mm}}
\vspace{-9truecm}
\caption{The HI surface density of DDO 47
(points) with the fit used in eq(6) (solid line). Units are
$M_{\odot}/{\rm pc}^2$ {\it vs} arcsec }
\label{HI}
\end{figure*}
%c
%%
\section{Model Independent Analysis} 

The high resolution VLA HI observations we use here are
discussed in detail in WB01. In summary, multi--array VLA observations
of DDO 47 give a resolution of $7.8'' \times 7.2''$ ($ 170 \  pc \times 140
\  pc $)  and a velocity resolution: 2.6 km/s$^{-1}$. The rotation curve, 
presented in WB01, is based on a moment 1 map of the HI data which has been
convoluted to $30''$ resolution and has been derived with the task ROTCUR in the
GIPSY package (see WB01 for details). For the new analysis of this paper, we have
refined the previously derived curve in WB01 by also subjecting the
data to the task INSPECTOR in GIPSY. The main results are essentially the
same, but now spatial and amplitude uncertainties are quantified.
The best-fit parameters are summarized here:
$v_{\rm sys}=272\ {\rm km/s}$; inclination i=35$^{\circ}$; position angle
PA=310$^{\circ}$ (see WB01 for details of the fitting procedure). The data points
presented in Fig. 1 are all independent measurements.

\begin{figure*}
\vspace{-4.3truecm}
\center{\psfig{file=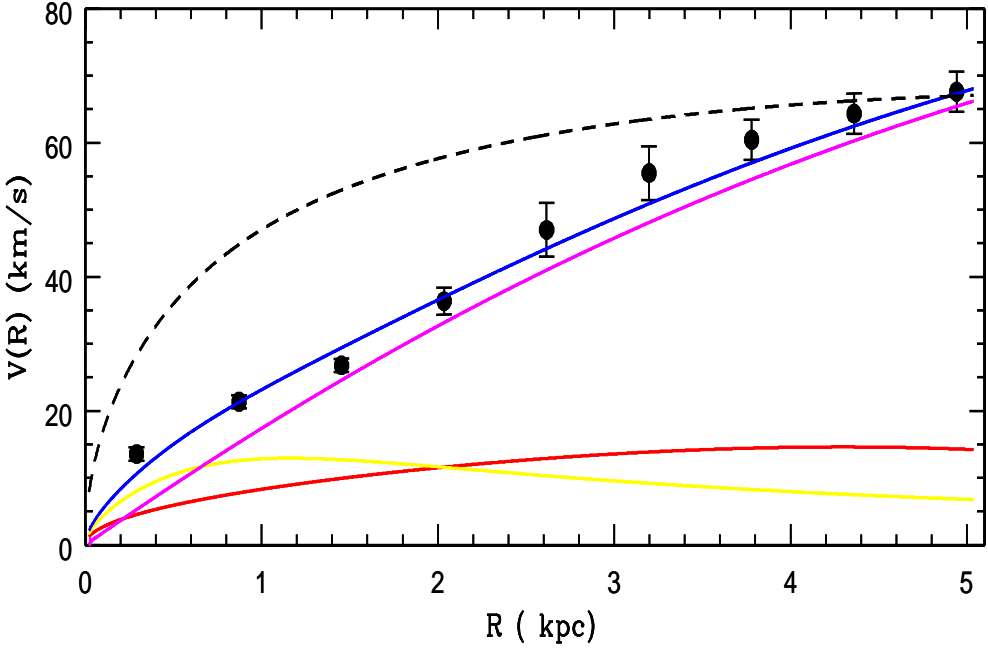,width=140mm,height=180mm}}
\vspace{-4.7truecm}
\caption{DDO 47 rotation curve reproduced by the BD2 model. Also shown the stellar (red), 
the gas  (yellow) and the Burkert halo distribution (magenta). Dashed line represents the 
$\Lambda$CDM mass model. }
\label{RC}
\end{figure*}

DDO 47's rotation curve (see Fig.\ref{RC}) increases with radius almost
{\it linearly}: we immediately infer that it is dominated by a
dark halo (Persic and Salucci, 1988, 1990) and then free from significant
baryonic contributions: these, at $r >(2-3) r_d$, would reveal themselves by causing a radial 
decrease or, at least, a marked deceleration in 
$V(r)$. The present data directly 
probe the 
gravitational potential of the 
dark halo inside  a volume about 50 times larger than the sphere 
enclosing the stellar disk.

Quantitatively, the circular velocity increases 
from 27 km/s at 1.5 kpc, corresponding to $2.7\ r_d$ (and $\sim 0.4 \ r_s$), 
up to 68 km/s at 5 kpc, (the farthest data point), corresponding to $9.1 \ 
r_d$ (and $\sim 1.3 r_s$).  In this region,  fitting linearly  the velocity data, 
 we derive the average  circular
 velocity slope: $ <dlog\  V/ dlog \ r> = 0.80 \pm 0.06$.
The CDM halo velocity slope, with an intrinsic upper limit of $0.5$ 
at $r=0$, predicts, for the same region, values  between $0.1$ and $0.3$.
Therefore, in DDO 47, over a radial excursions of 6 disk scale--lengths,   we detect an  {\it 
increase}  of 600\%, in $V^2$,  that results  about one order of magnitude larger than that  
expected for a NFW halo.

A good estimate of the {\it total} galaxy density $\rho_{tot}$ is given, in a model 
independent way, by:

\begin{equation}
\rho_{tot}(r) = \frac{V(r)^2}{4 \pi\ G\ r^2 } \Big(1+2\ \frac{d\log \ V(r)}{d\log\  r} \Big) 
\end{equation}

at $1\  kpc$ we get:   $\rho_{tot}(1) \simeq 2 \times 10^{-24}\  g/cm^3$, that we compare, by 
means 
of eq 
(1) and (3),  with the  density of   $\Lambda CDM$ halos  in the mass range of  $log \  
M_{vir}/ M_\odot=10-11$; in all cases: $\rho_{ \Lambda CDM}(1\ kpc) \sim 10^{-23}$, i.e. a value 
well exceeding that observed. This  normalization of  the density profile 
(1)  
implies that NFW dark halos  dominate the mass distribution   
of galaxies  with  baryonic  distributions like DDO 47  in a more prominent way than cored  halos 
would do.   
   
Thus,  the CDM crisis is not only the "wrong" radial density  distribution of 
 the halos around galaxies, but especially  the fact that, while we easily
detect halos with a "core", we fail to detect halos with  NFW "cusps", despite that in  
$\Lambda$CDM scenario the latter should be, with respect to the luminous matter, much more 
prominent.

\section{Mass Modeling} 
The DDO~47 baryonic components are:

1) Exponential thin disk of stars of scale--length $r_d=28"\pm 2"$, (0.54 kpc), Makarova et al., 
(2001) contributing to the circular velocity $V$  as: ($y \equiv r/r_d$) 

\begin{equation}
V^2_d(r)=1.28\ \beta\ V^2(R_{opt})\ y^2\ (I_0K_0-I_1K_1)|_{1.6 y}
\label{Vd}
\end{equation}

$\beta$ is the disk velocity fraction at $R_{opt} \equiv 3.2\ r_d$ (for solid-body rotation curves 
$\beta \sim 0.1$, Persic and Salucci, 1990).

2) HI disk: the HI surface density is shown in Fig.\ref{HI}, the gaseous disk total mass,  
including  the He contribution (30 \% of the HI mass), is  about $3  \times 
10^8 M_{\odot}$; the (HI+He) contribution to the circular velocity is:
 
\begin{equation}
V^2_g (r)= (9.2 \ r ^{0.49} -1.6\times 10^{-2} r ^{3.5}) D_4^2
\end{equation}

where $D_4$ is the galaxy distance in units of the reference distance of 4 Mpc, 
likely ranging between 0.8 and 1.2. For 
$r>3 \ kpc $, $V_g(r)$ is approximately constant,  unlike the circular velocity.

We model the mass distribution of DDO 47 with 3 components: a dark halo with a   Burkert profile,
a stellar   disk and gaseous disk.The aim of this  mass model (hereafter BD2) is to reproduce 
the available kinematical  data, leaving to future work to derive a "global"  mass model  that 
can be  extrapolated outside the range in which  data are available.  The density, for a Burkert 
halo, is:

\begin{equation}
\rho_{B}(r) = \frac{\rho_0\ r^3_0}{(r+r_0)(r^2+r_0^2)}
\label{rhoBBS}
\end{equation}

where $\rho_0$ and $r_0$ are the central density and the size of the region of (almost) constant 
density. Then:
 $M_{\rm B}(r) = 4\ M_0\ \{ \ln (1 + r/r_0) -\arctan (r/r_0) + 0.5 \ln [1
+(r/r_0)^2]\}
$
with $M_0 = 1.6\ \rho_0 r_0^3 $ Then: $V^2_B(r) =G\ M_B(r)/r$.
The BD2 model has 2 (really)  free parameters  $\rho_0$, $r_0$ and two  constrained "free"
parameters: $D_4$ and $\beta$. They  are derived by 
$\chi$--square minimization of the quantity
$V^2(r)-V^2_d(r)-V^2_B(r)-V^2_{g}(r) $ over the 9 data points.
We find $\rho_0=(1.4 \pm 0.4)\times 10^{-24}
g/cm^3$, $r_0 = 7^{+2}_ {-1} $ kpc, $\beta= 0.13\pm 0.04$, $D_4=0.9\pm 0.1$.
The best--fit mass model
is shown in Fig.\ref{RC}: the data are reproduced in an excellent way.
The total disk mass  results:  $M_d = (5 \pm 1) \times 10^7
M_{\odot} $ and  the halo mass inside 5 kpc $M_B(5)=5 \times 10^9 M_{\odot}$
Since the RC does not reach the region where $V_B$ starts to 
decelerate (i.e. when $r >r_0$),   we cannot extrapolate 
 the  (dark)  mass distribution outside the region directly mapped by data.

It is illustrative to show the "best" mass model for the $\Lambda$CDM scenario.
We know a-priori that it will not be successful in representing the data because NFW halos have 
a density profile intrinsically inconsistent with that of the dark halo around DDO 47. It is 
worth, however to qualify the discrepancy: in Fig (2) we report the $\Lambda$CDM mass model 
with the dashed line, $V_{\Lambda CDM}$ has only a free parameter, the virial mass that we fix by 
matching 
 the model with  the outermost velocity data: $M_{vir}=4\times 10^{10} M_\odot$. 
Notice that there are not appreciable changes by including the gas and 
the stellar disk contributions. 
The discrepancy of the CDM velocity model with the  ever--rising RC
data (see Fig.\ref{RC}) needs not commenting; it is however instructive to look at one particular 
aspect of the failure of $\Lambda$CDM: let us derive
its density at $r=5 \ kpc$, the radius where we matched model and data. From eq 
(1)  we get a value of $2 \times 10^{-25} g/cm^3$, {\it smaller} than the corresponding 
model-independent one. 
CDM theory shows, then, a global inconsistency with data which is well beyond a local 
disagreement: 
we can describe it as CDM bringing in the ``cusp" region much more dark mass than expected, while 
leaving the region outside the ``cusp" with a mass deficit.

We conclude this section by stressing the main  result  obtained: 
we start from the rising circular velocity of  DDO 47:  $ <dlog \ V/ dlog\  r> =0.80 \pm 0.06$   
and we   find that,  out to 9 disk-scalenghts, the slope of the  dark halo  contribution to $V$,  
$dlog \  V_h/ dlog \  r=(dlog \  \rho/dlog \ r +2)/2$,  is steeper: $ dlog \  V_h/ dlog \ r = 0.90 
\pm 
0.03$ (due to small baryonic components that induce  a   
slight deceleration in    the rotation curve).  On the other side, 
 $\Lambda$CDM  mass 
modeling of DDO 47,    in the same region, yields to an unacceptably shallower halo profile:  $ 
dlog \  V_{\Lambda CDM} =0.2\pm 0.1$. Thus,  mass modeling widens the model-independent 
discrepancy between  the dark halo around 
DDO 47 and   $\Lambda$CDM  NFW halos discussed in the previous section.

%________________________________________________________________

\section{Conclusions: an Intriguing Evidence}

We think that $\Lambda$CDM NFW halos  have  lost, in DDO 47, the last call to represent the dark 
matter   around 
galaxies. This galaxy, in fact, 
in terms of extension, spatial resolution, regularity and smallness of the observational errors of 
the rotation curve and  in terms of the large dark-to-luminous mass and $r_s$/(spatial resolution 
of rotation curve)  ratios,   is a perfect laboratory  to detect a NFW halo  by pin-pointing the 
density slope change, from 0.5 to about  0,  that should occur exactly in the region mapped by 
data.   
 Saying it plainly,  DM halo density of DDO 47, out to $9\ r_d$, is instead  fully and uniquely
determined by two parameters, a  core density and a core
radius,   that are {\it not even existing} in the
gravitational instability/hierarchical clustering Cold Dark Matter
scenario.

To reconcile the  $\Lambda$CDM theory with this evidence is clearly beyond the scope of this 
paper, let us however  just indicate two possible routes:
 1- the (gravitational) physics of  the collapse of  the innermost 10\%  of the halo mass 
distribution could be  more complex than that modeled by  current CDM simulations.
 2- a (yet)
unknown physical process could occur in the innermost $10^{-3}$\% of
the dark halo volume, cutting down the post--collapse DM density by
$1-2$ orders of magnitudes.

%________________________________________________________________

\section{Acknowledgments}

We want to thank the referee R. Bottema for suggestions that have  improved    
the presentation of this paper. P.S. wants to thank G. Danese and U. Klein  for useful discussions.

%________________________________________________________________

\end{document}